\documentclass[aps,amsmath,amssymb,prb,twocolumn,showpacs,floatfix]
{revtex4}
\usepackage{bm}
\usepackage{epsfig}

\begin{document}

\title{A Scaling Approach to Ideal Quantum  Gases}

\author{Thomas Nattermann\footnote{natter@thp.uni-koeln.de}}
\affiliation{Institut f\"ur Theoretische Physik, Universit\"at zu
K\"oln, 50937 K\"oln, Germany}

\date{\today}
\begin{abstract}
The  thermodynamic properties of ideal quantum gases are derived
solely from dimensional arguments, the Pauli principle and
thermodynamic relations, without resorting to statistical mechanics.
\end{abstract}

\maketitle

\newcommand{\Lr}{\lambdabar_{T,1}}
\newcommand{\Ln}{\lambdabar_{T,2}}
\newcommand{\Lc}{\lambdabar_{C}}

\renewcommand{\baselinestretch}{0.7}

%%%%%%%%%%%%%%%%%%%%%%%%%%%%%%%%%%%%%%%%%%%%%%%%%%%%%%%%%%%%%
%%%%%%%%%%%%

\section{\textsc{Introduction}}\label{Introduction}

%%%%%%%%%%%%%%%%%%%%%%%%%%%%%%%%%%%%%%%%%%%%%%%%%%%%%%%%%%%%%%
%%%%%%%%%%%%

In this article we present a derivation of the thermal equations of
state of ideal quantum gases by using solely dimensional analysis,
thermodynamic considerations and simple physical arguments
\cite{notnew}. The thermodynamic relations of gases depend strongly
on the energy-momentum relation of the gas particles and on their
spin, i.e., whether the particles  are fermions or bosons.
 The energy-momentum
relation of free massive particles is given by
\begin{equation}\label{eq:relativistic_dispersion_relation}
    E_{\bf p}=\sqrt{m^2c^4+c^2{\bf p}^2}-mc^2.
\end{equation}
$m$  denotes the rest mass of the particles, $c$ is the velocity of
light. Here we have subtracted the rest energy $mc^2$ from the
particle energy which results in a redefinition of the chemical
potential, i.e. $\mu$ has to be replaced  by $\mu-mc^2$ further on.
In the \emph{non}-\emph{relativistic} limit $mc\gg |{\bf p}|$,
$E_{\bf p}\approx \frac{{\bf p}^2}{2m}$, whereas in the
\emph{ultra}-\emph{relativistic} limit $mc\ll|{\bf p}|$, $E_{\bf
p}\approx c|{\bf p}|$. Both cases can be written in the single
exponent dispersion relation
   \begin{equation}
   E_{\bf p}\sim mc^2\left(\frac{|{\bf
   p}|}{mc}\right)^{\nu}\sim \gamma_{\nu}\,\,|{\bf
   p}|^{\nu},\,\,\,\,\,\,
   \gamma_{\nu}=m^{1-\nu}c^{2-\nu},
   \label{eq:dispersion_relation}
   \end{equation}
where $\nu=1,2$ for relativistic and non-relativistic particles,
respectively.

Another group of particles to which our considerations apply are
\emph{quasi}-\emph{particles} in condensed matter systems.
 In these  systems deviations from the state of perfect
order can be described by a superposition of quantized elementary
excitations which behave as bosonic {particles} with an energy
momentum relation also of the form (\ref{eq:dispersion_relation}).
The quasi-particles we are considering,  as well as photons, have a
vanishing chemical potential, i.e. their number is not fixed. At low
temperatures this number  is small and hence their interaction can
be neglected. Quasi-particles are characterized by a fixed value of
$\nu>0$\,\,\, \cite{quasiparticles}. $\gamma_{\nu}$ has to be
expressed in terms of the quasi-particle parameters. Examples for
the case $\nu=1$ are lattice vibration of solids with acoustic
phonons
 as quasi-particles \cite{LLV1} ($c$ denotes now the sound velocity)
or spin waves in anti-ferromagnets  with magnons as
quasi-particles \cite{LLV2}  ($c$ denotes here the spin wave
velocity). An example with $\nu=2$ are magnons in ferromagnets
\cite{LLV3}, where $m\sim \hbar^2/(b^2T_c)$.  $b$ denotes the
lattice constant and $T_c$ the Curie temperature. For further
examples of quasi-particles see e.g. reference \cite{Anderson}.

 It is interesting to remark that our considerations can
be  extended to \emph{non}-\emph{integer} values of $\nu$. One
example are capillary wave excitations
 of the surface of superfluid $^4He$, the so-called ripplons
 \cite{Lederer} which
 show a dispersion relation (\ref{eq:dispersion_relation}) with
 $m/c\sim {\rho\hbar/\sigma}$ and $\nu={3/2}$. Here $\sigma$
 and $\rho$
 denote surface tension and the liquid density, respectively.
  Our results apply also to this case,
 but in the further discussion we will mainly focus on the
 cases of integer
 values of
 $\nu$.

%%%%%%%%%%%%%%%%%%%%%%%%%%%%%%%%%%%%%%%%%%%%%%%%%%%%%%%%%%%%%%%%%%%%%%

\section{\textsc{Dimensional analysis }}

\subsection{The classical case}

%%%%%%%%%%%%%%%%%%%%%%%%%%%%%%%%%%%%%%%%%%%%%%%%%%%%%%%%%%%%%%
%%%%%%%%%%

 Considering
a homogeneous gas of $N$ point particles confined in a container of
volume $V=vN$ and coupled to a bath of temperature $T$, the free
energy can be written in the form
   \begin{equation}
   F(T,V,N)=NF(T,v,1)=Nf(T,v)\,.
   \label{eq:free_energy}
   \end{equation}
$f(T,v)$ is the free energy per particle. The mean separation $a$
between particles is related to the volume per particle  $v=a^d$,
where $d$ denotes the dimension of the system. The pressure follows
then from $p=-({\partial f}/{\partial v})_{T}$.

We begin with the consideration of a \emph{classical gas} with fixed
exponent $\nu$ of the dispersion relation
(\ref{eq:dispersion_relation}). In this case the pressure can only
depend on the three parameters $\gamma_{\nu}$, $a$ and $T$. There is
no further parameter. The temperature $T$ appears in thermodynamics
in the relation $T^{-1}= \partial S/\partial E$, where $S$ and $E$
denote the entropy  and the energy, respectively. Entropy is
determined only up to a multiplicative (Boltzmann) constant
\cite{Lieb}.  For the moment we will only assume that this constant
is dimensionless, i.e. we measure $T$ in energy units.

Denoting the dimension of mass, length and time by $M$, $L$ and $t$,
respectively, the dimension of $\gamma_{\nu}$, $a$, $T$ and $p$ are
$M^{1-\nu}L^{2-\nu}t^{\nu-2}$,  $L$, $ML^2t^{-2}$ and
$ML^{2-d}t^{-2}$, respectively. Since the three parameters have
different dimensions,  there is a  unique combination of these which
has the dimension of  pressure \cite{dimensional_analysis}
   \begin{equation}
   p\sim\frac{T}{a^d},\,\,\,\,\,\,\,\,\,\,\,\,\,\, a^d=v=\frac{V}{N},
   \label{eq:p_classical}
   \end{equation}
as expected. We  fix now our temperature scale  by choosing the
constant of proportionality in (\ref{eq:p_classical}) to be equal
one, i.e. $pv=T$, which is the known result
\cite{Boltzmann_constant}.

This suggest $f(T,v)\sim T\ln v$. Clearly $v$ should occur in the
free energy in combination with another volume to make the argument
of the logarithm dimensionless. This shows the limitations of
classical physics.

In the case of massive particles there is only a dependence on $m$
\emph{or} $c$ as long as we use the single power law relation
(\ref{eq:dispersion_relation}) with $\nu= 1 $ {or} $2 $ fixed.
However, for the description of the \emph{cross}-\emph{over} between
the non-relativistic and the relativistic regime  both $m$
\emph{and} $c$ occur and hence a new dimensionless parameter
$T/(mc^2)$ appears. The relation (\ref{eq:p_classical}) is formally
correct even in the region $T\gg mc^2$. However, one has to expect
that in this region pair creation processes will take place and a
grand canonical description is more appropriate, as we will consider
in section \ref{Grand_Canonical_Description}.

In the case of \emph{charged} particles  the ideal gas properties
are only preserved under certain conditions. If we consider for
simplicity electrons (on a neutralizing positively charged
background)
 the electron-electron interaction is of the order $e^2/a$.
To have still an ideal gas the Coulomb energy has to be small
compared with the thermal energy, i.e. $e^2/a\ll T$ or  $
a_T=e^2/T\ll a$ \quad  \cite{aT}.

%%%%%%%%%%%%%%%%%%%%%%%%%%%%%%%%%%%%%%%%%%%%%%%%%%%%%%%%%%%%%%%%%%%%

\subsection{Quantum case}\label{Quantum_case}

%%%%%%%%%%%%%%%%%%%%%%%%%%%%%%%%%%%%%%%%%%%%%%%%%%%%%%%%%%%%%%%%%%%%

In the {quantum case} the Planck constant of action $\hbar$ appears
as a new parameter. The pressure is now no longer uniquely fixed by
dimensional analysis since in order  to determine the four exponents
of $\gamma_{\nu}, a, T$ and $\hbar$ only three relations (from the
comparison of the powers of $t$, $L$ and $M$) exist. Thus the
equation of state will  depend additionally on a
\emph{dimensionsless parameter} $x_{\nu}$ which includes $\hbar$ as
well as $T$, $a$ and $\gamma_{\nu}$. Dimensional analysis shows that
the only possible choice is \cite{powers}
   \begin{equation}
   x_{\nu}=\frac{a}{\hbar}\left(\frac{T}{\gamma_{\nu}}\right)^{1/\nu}
    =: \frac{a}{\lambdabar_{T,\nu}}\,\,\,,
   \label{eq:x}
   \end{equation}
\begin{equation}\label{}
\lambdabar_{T,\nu} =
\hbar\left(\frac{{\gamma_{\nu}}}{T}\right)^{1/\nu}=
\lambdabar_C\left(\frac{mc^2}{T}\right)^{1/\nu},\,\,\,
\lambdabar_C=\frac{\hbar}{mc}.
\end{equation}
Here $\lambdabar_{T,\nu}$ denotes the {thermal} {de} {Broglie}
{wave} {length} of the (quasi-) particles.  $\lambdabar_C$ is the
{Compton} {wave} length in the case of massive particles. We ignore
all numerical pre-factors since they are beyond the accuracy of the
dimensional analysis presented here. Since for $\hbar \to 0$, i.e.
$\lambdabar_{T,\nu}\ll a$, we have to regain the classical result
(\ref{eq:p_classical}), quantum behavior will be seen for $x_{\nu}
\lesssim 1$, i.e. for $a\lesssim\lambdabar_{T,\nu}$. The borderline
$x_{\nu}\approx 1$ defines a \emph{characteristic} \emph{energy}
$\gamma_{\nu}(\hbar/a)^{\nu}$ which turns out to be the Fermi energy
for fermions and the temperature of Bose condensation for bosons,
respectively.
 This
will be discussed in detail in the subsequent section.

\begin{figure}[hbt]
   \centerline{\epsfxsize=6cm
   \epsfbox{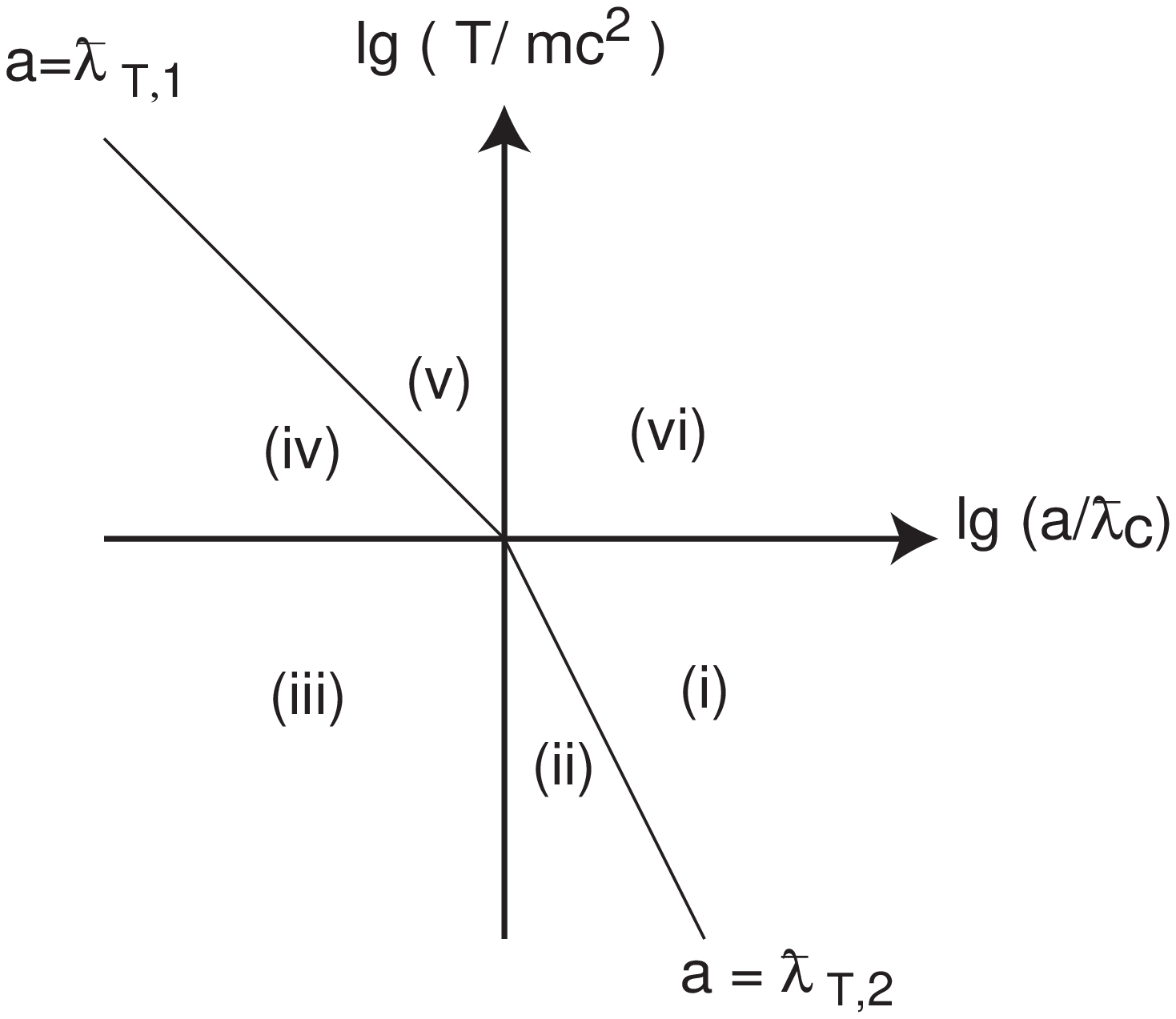}}
   \caption{\label{fig:idal_gas1} The various regions (i) - (iv)
   of the parameter space of massive particles depicted in
   the $\text{log}(T/mc^2)$ versus
   $\text{log}(a/\Lc)$ plane.
Massive matter under ordinary conditions
   ($T\ll mc^2$)
   is located in a region well below  the horizontal axis.
   For more explanations see the text.}
   \end{figure}
In addition to the  cross-over from the classical to the quantum
regime, \emph{massive particles} also show a {cross}-{over} from
non-relativistic to relativistic behavior.  If the mean particle
distance is given by $a$, then for $a\gg \Lc$, the typical particle
energy $E_{|{\bf p}|\sim \hbar/a}$ approaches the non-relativistic
expression $\hbar^2/(ma^2)$ whereas for $a\ll \Lc$ the relativistic
expression $c\hbar/a$ applies.

From the mutual order of the three length scales $a, \Lc$ and
$\lambdabar_{T,{\nu}}$ \emph{six} regions can be distinguished
\cite{lambda}. Anticipating the result of the detailed discussions
of  the following sections these six regions exhibit very different
physical properties :
\begin{description}
    \item[(i)\,\,\,\, $\Lc<\lambdabar_{T,2}<a$]:
    This is the region of {{classical}}
    behavior. The only relevant length scale
    is the the inter-particle distance $a$.
    \item[(ii) \,\, $\Lc<a<\Ln$]: The region of {{non}-{relativistic}
    {quantum}} behavior. Fermions are degenerated, bosons
    condense. Since
    $\Lc<a,\lambdabar_{T,\nu}$ relativistic effects do not play a role.
    \item[(iii)\,\,  $a<\Lc<\lambdabar_{T,2}$]: A region of quantum
    behavior where
     bosons behave non-relativistically (as in (iii)) whereas fermions
     exhibit
     relativistic effects. The Pauli principle makes the difference.
    \item[(iv) \,\, $a<\Lr<\Lc$] : Fermions show still
    low temperature
    behavior (as in
    (iii)), bosons exhibit particle-antiparticle pair
    production.
     %(as well as  a condensate).
    \item[(v) \,\, $\Lr<a<\Lc$] : Fermions and bosons are described by
     the same
    equation of state as photons.% the same as
     \item[(vi)\,\,  $\Lr<\Lc<a$] : Same behavior as in (v).
\end{description}

Bose condensation as well as   particle-antiparticle pair production
will in general  change the  effective inter-particle distance from
$a$ to $\Ln$ and $\Lr$, respectively. Since pair production keeps
some of the quantum numbers constant (e.g. the total charge or the
lepton or hadron number) in these regions $a$ is a measure of the
corresponding quantum number of the gas.

\medskip

For \emph{charged} fermions dimensional analysis gives a further
intrinsic length scale, the Bohr radius
$a_{\text{Bohr}}=\hbar^2/(me^2)$. For  simplicity we consider again
a gas of electrons on a positively charged background. To have still
an ideal gas the Coulomb interaction $\sim e^2/a$ has to be small
compared to the kinetic energy. For non-relativistic electrons this
results in $e^2/a\ll
 \hbar^2/(ma^2)$, i.e., $a\ll \hbar^2/(me^2)=a_{\text{Bohr}}$.
 For relativistic electrons one obtains in the same way $e^2/a\ll
 \hbar c/a$ or $e^2/(\hbar c)\approx 1/137\ll 1$, i.e., the ideal
 gas condition is always fulfilled.
For non-relativistic charged bosons outside the condensate, $a$ has
to be replaced
 by
 $\lambdabar_{T,2}$ (see below)
 which leads to the  condition $\Ln<a_{\text{Bohr}}$. The effect
 of Coulomb interaction will be more pronounced in the condensate
 since the kinetic energy is zero. Indeed, it has been argued that
 there is
 \emph{no} Bose condensation in the presence of Coulomb interaction
 \cite{Lee}.

\medskip

The different regions of the parameter space can be visualized in
the diagram depicted in Figure \ref{fig:idal_gas1} where we plot $
{T}/{mc^2}=(\Lc/\lambdabar_{T,\nu})^{\nu}$ versus
${a}/{\lambdabar_C}$ in a double logarithmic representation. The
lines $T=mc^2$ and $a=\Lc$ correspond to the horizontal and
vertical axis, respectively.  The  line $a=\lambdabar_{T,\nu}$
separates the classical region $a>\lambdabar_{T,\nu}$ from the
quantum region $a<\lambdabar_{T,\nu}$\,\,. In our double
logarithmic plot this line is straight with a slope equal to
$-\nu$.
 {Ordinary } massive matter at temperatures
realized on earth is restricted to the region below the horizontal
axis. For electrons $m_e\,c^2\sim 10^{10} K$, and hence for room
temperatures $T\sim 10^2 K$ this region is given by $
{T}/{mc^2}\le 10^{-8}$. The Bohr radius $a_B$ corresponds to
${a_B}/{\Lc}= {\hbar c}/{e^2}\approx 137$.

%%%%%%%%%%%%%%%%%%%%%%%%%%%%%%%%%%%%%%%%%%%%%%%%%%%%%%%%%%%%%%%%%%

\subsection{Grand Canonical Description}
\label{Grand_Canonical_Description}

%%%%%%%%%%%%%%%%%%%%%%%%%%%%%%%%%%%%%%%%%%%%%%%%%%%%%%%%%%%%%%%%%%%%

So far we assumed that the particle number $N=V/a^d$ and hence $a$
was fixed. In some cases it is more appropriate to consider the
{grand} {canonical} description where instead of  $N$ the
\emph{chemical} \emph{potential} $\mu$ is given.  This description
is in particular useful in those parameter regions where particle
creation and annihilation processes (like the inverse $\beta$-decay
or electron-positron pair creation) become important.

The corresponding thermodynamic potential $J(T,V,\mu)$ is related to
the pressure by the Gibbs-Duham relation
\begin{equation}\label{eq:J}
 J(T,V,\mu)=VJ(T,1,\mu)\equiv V j(T,\mu)  =-pV.
\end{equation}
Since $T$ and $\mu$ have the same dimension there is \emph{no}
intrinsic length scale and hence no purely classical expression
for the pressure. If we include $\hbar$ as a further parameter
\begin{equation}\label{eq:lambda_grand_canonical}
\lambdabar_{T,\nu}\,\,\omega\left(\frac{\mu}{T}\right)
\end{equation}
appears as the \emph{only} length scale of the problem.
$\omega\left(\frac{\mu}{T}\right)$ is so far an arbitrary but
dimensionless function of $\mu/T$. Following our previous
dimensional arguments the pressure can now be written as
\cite{LLV1_56_9,LLV1_61_7}
\begin{equation}\label{eq:p_grand_canonical}
    p=\frac{T}{\lambdabar_{T,\nu}^d}\Omega\left(\frac{\mu}{T}\right).
\end{equation}
with $\Omega\left(\frac{\mu}{T}\right)$ likewise so far unknown. For
particles with vanishing chemical potential the pressure is
\emph{always} given by equation (\ref{eq:p_grand_canonical}) with
$\Omega(0)\sim 1$. Important examples are photons, magnons in
anti-ferromagnets and acoustical phonons (all $\nu=1$), magnons in
ferromagnets ($\nu=2$) and ripplons ($\nu=3/2$).

\medskip

With (\ref{eq:J}) we get
$J=-{T}{}\Omega(\frac{\mu}{T})V\lambdabar_{T,\nu}^{-d}$ and hence
\begin{equation}\label{eq:N}
\frac{\partial J}{\partial \mu}=-N
=-\frac{V}{\lambdabar_{T,\nu}^d}\,\Omega'\left(\frac{\mu}{T}\right)
=-\frac{pV}{T}\,\frac{\Omega'}{\Omega}
\end{equation}
Using the classical equation of state (\ref{eq:p_classical}) this
requires $ d{\Omega}_{cl}/dx={\Omega}_{cl}$. Hence for a classical
gas
\begin{equation}\label{eq:Omega_of_mu/T}
\Omega_{cl}\left(\frac{\mu}{T}\right)=\Omega_{cl} (0)\,e^{{\mu}/ T}.
\end{equation}
(\ref{eq:p_grand_canonical}) and (\ref{eq:Omega_of_mu/T}) can be
used to calculate the chemical potential of the classical gas
\cite{LLV1_45_5}
\begin{equation}\label{eq:chemical_potential}
    \mu=T
    \ln\left(\frac{p\lambdabar_{T,\nu}^d}{T\Omega_{cl}(0)}\right)=T
    \ln\left(\frac{\lambdabar_{T,\nu}^d}{a^d\Omega_{cl}(0)}\right).
\end{equation}

%%%%%%%%%%%%%%%%%%%%%%%%%%%%%%%%%%%%%%%%%%%%%%%%%%%%%%%%%%%%%%%%%%%%%%

\section{\textsc{Equation of state }}

%%%%%%%%%%%%%%%%%%%%%%%%%%%%%%%%%%%%%%%%%%%%%%%%%%%%%%%%%%%%%%%%%%%%%
%%%%

\subsection{Weak quantum corrections}

%%%%%%%%%%%%%%%%%%%%%%%%%%%%%%%%%%%%%%%%%%%%%%%%%%%%%%%%%%%%%%%%%%%
%%%%%%%

 Since for $x_{\nu}\to\infty$ the system has to approach
the classical limit we make the following Ansatz for the pressure
   \begin{equation}
   p=\frac{T}{a^d}\psi_{\text {F/B}}(x_{\nu})
   \label{eq:p_scaling_ansatz}
   \end{equation}
with $\psi_{\text {F/B}}(x_{\nu}\to\infty)\to\text{const}$. Here the
subscript F or B stands for fermions or bosons. The scaling function
$\psi_{\text {F/B}}(x_{\nu})$ describes the cross-over from
classical to quantum behavior of particles with a dispersion
relation with fixed value of $\nu$. For massive particles this
Ansatz can be generalized to
\begin{equation}
   p=\frac{T}{a^d}\Psi_{\text{F/B}}\left(x_1,\frac{a}{\Lc},\right)
   \label{eq:p_scaling_ansatz_extended}
   \end{equation}
which describes in addition also the cross-over from
non-relativistic to ultra-relativistic behavior. The scaling
function $\Psi_{\text{F/B}}(x_1,y)$ can be related to
(\ref{eq:p_scaling_ansatz}) in the limiting cases $y\to \infty$ and
$y\to 0$, respectively.  In the non-relativistic limit $a\gg\Lc$,
i.e. $y\to \infty$, $\Psi_{\text{F/B}} (x_1,y)\sim
\psi_{\text{F/B}}(\sqrt {x_1y})=\psi_{\text{F/B}}(x_2)$. In the
ultra-relativistic case $a\ll \Lc$, i.e. $y\to 0$,
$\Psi_{\text{F/B}} (x_1,y)\sim \psi_{\text{F/B}}(x_1)$. In the
following we will mainly concentrate on the discussion of
$\psi_{\text{F/B}}(x_{\nu})$.

We consider first the case of \emph{weak} {\emph{quantum}}
\emph{fluctuations}. Assuming that the pressure can be expanded in
powers of the particle density, i.e. in powers of $x_{\nu}^{-d}$,
(the classical result (\ref{eq:p_classical}) is the lowest order
term of this expansion) $\psi_{\text{F/B}}(x_{\nu})$ can be written
in the form
   \begin{equation}
   \psi_{F/B}(x_{\nu})\approx 1\pm\text{const.}
   \,x_{\nu}^{-d}\,,\quad
   x_{\nu}\gg 1\,.
   \label{eq:p_weak_corrections}
   \end{equation}
Here the plus and minus sign corresponds to fermions and bosons,
respectively. Indeed, for fermions the Pauli principle will lead
to an increase of the pressure  with respect to the classical
case. For bosons the pressure is reduced since quantum mechanics
increases the probability for the double occupancy of a state
\cite{probability}. (\ref{eq:x}), (\ref{eq:p_scaling_ansatz}) and
(\ref{eq:p_weak_corrections}) describe indeed the lowest order
quantum correction to the classical equation of
state\cite{ll55.15}.

%%%%%%%%%%%%%%%%%%%%%%%%%%%%%%%%%%%%%%%%%%%%%%%%%%%%%%%%%%%%%%%%%%%%
%%%%

\subsection {\textsc{Strong quantum limit - fermions}}

%%%%%%%%%%%%%%%%%%%%%%%%%%%%%%%%%%%%%%%%%%%%%%%%%%%%%%%%%%%%%%%%%%%
%%%%%%

In the \emph{strong quantum limit} $x_{\nu}\to 0$, we have  even
more pronounced differences between fermions and bosons. For
\emph{fermions} the Pauli-principle guarantees a non-zero pressure
even for $T\to 0$. From (\ref{eq:x}) and
(\ref{eq:p_scaling_ansatz}) this requires $\psi_F(x_{\nu})\sim
x_{\nu}^{-\nu}$ and hence
\begin{equation}\label{eq:p_fermions_general}
    p_F(\nu)\sim \frac{\hbar^{\nu}\gamma_{\nu}}{a^{d+\nu}}\,
    \sim \frac{E_{F,\nu}(a)}{a^d}.
\end{equation}
On the r.h.s. we have introduced the \emph{Fermi} {\emph{energy}}
\begin{equation}\label{eq:fermi_energy}
E_{F,\nu}(a)\sim   \frac{\hbar^{\nu}\gamma_{\nu}}{a^{\nu}}
%mc^2\left(\frac{\lambdabacal E}_{F,\nu}(a)}{a^d}. r_C}{a}\right)^\nu,
\end{equation}
which is the kinetic energy of the fermions with momentum $|{\bf
p}|\sim \hbar/a$. Apparently, in (\ref{eq:p_fermions_general}) the
Fermi energy replaces  the temperatures of the classical expression
(\ref{eq:p_classical}).
 In particular this gives for relativistic particles
\cite{ll614}
   \begin{equation}
   p_{\text{F}}(\nu=1)\sim\frac{\hbar c}{a^{d+1}},
   \label{eq:p_fermions_relativistic}
   \end{equation}
and for non-relativistic particles \cite{ll575}
   \begin{equation}
   p_{\text{F}}(\nu=2)\sim\frac{\hbar^2}{ma^{d+2}},
   \label{eq:p_fermions_non-relativistic}
   \end{equation}
respectively. Both cases can be combined in the scaling formula
   \begin{equation}
   p_{\text{F}}\sim\frac{\hbar c}{a^{d+1}}\,
   \,\frac{\Psi_{\text{F}}(x_1,\frac{a}{\Lc})}{\psi_F(x_1)}.
   \label{eq:p_fermions_crossover}
   \end{equation}
The precise form of the scaling function $\Psi_{\text{F}}(x,y)$
remains unknown in this approach. The cross-over between
(\ref{eq:p_fermions_non-relativistic}) and
(\ref{eq:p_fermions_relativistic}) takes place at $a\sim \Lc$, as
follows from the general properties of $\Psi_{\text{F}}(x,y)$
discussed before.

As long as $\Lc<a<\Ln$ (region (ii) in Figure 1) the particles show
 non-relativistic quantum behavior. This situation exists e.g.
  for electrons in solids ($a\sim
10^{-10}\,{\rm m}$, $\lambdabar_C\sim 10^{-13}\,{\rm m}$) or
neutrons in neutron stars ($a\sim 10^{-13}\,{\rm m}$,
$\lambdabar_C\sim 10^{-16}\,{\rm m}$).

On the other hand, for $a<\lambdabar_C$ the kinetic energy is larger
than $mc^2$ and the particles behave relativistically. One has to
expect that this 'low temperature' relativistic quantum behavior is
seen as long as the temperature is small compared to the Fermi
energy $E_{F,\nu}(a)\sim c\hbar/a$, i.e. for $a\lesssim \Lr $ (
corresponding to regions (iii) and (iv) of Figure 1). Relativistic
electrons exist e.g. in white dwarfs ($a\sim 10^{-13}\,{\rm m}$).
  \begin{figure}[hbt]
   \centerline{\epsfxsize=9cm
   \epsfbox{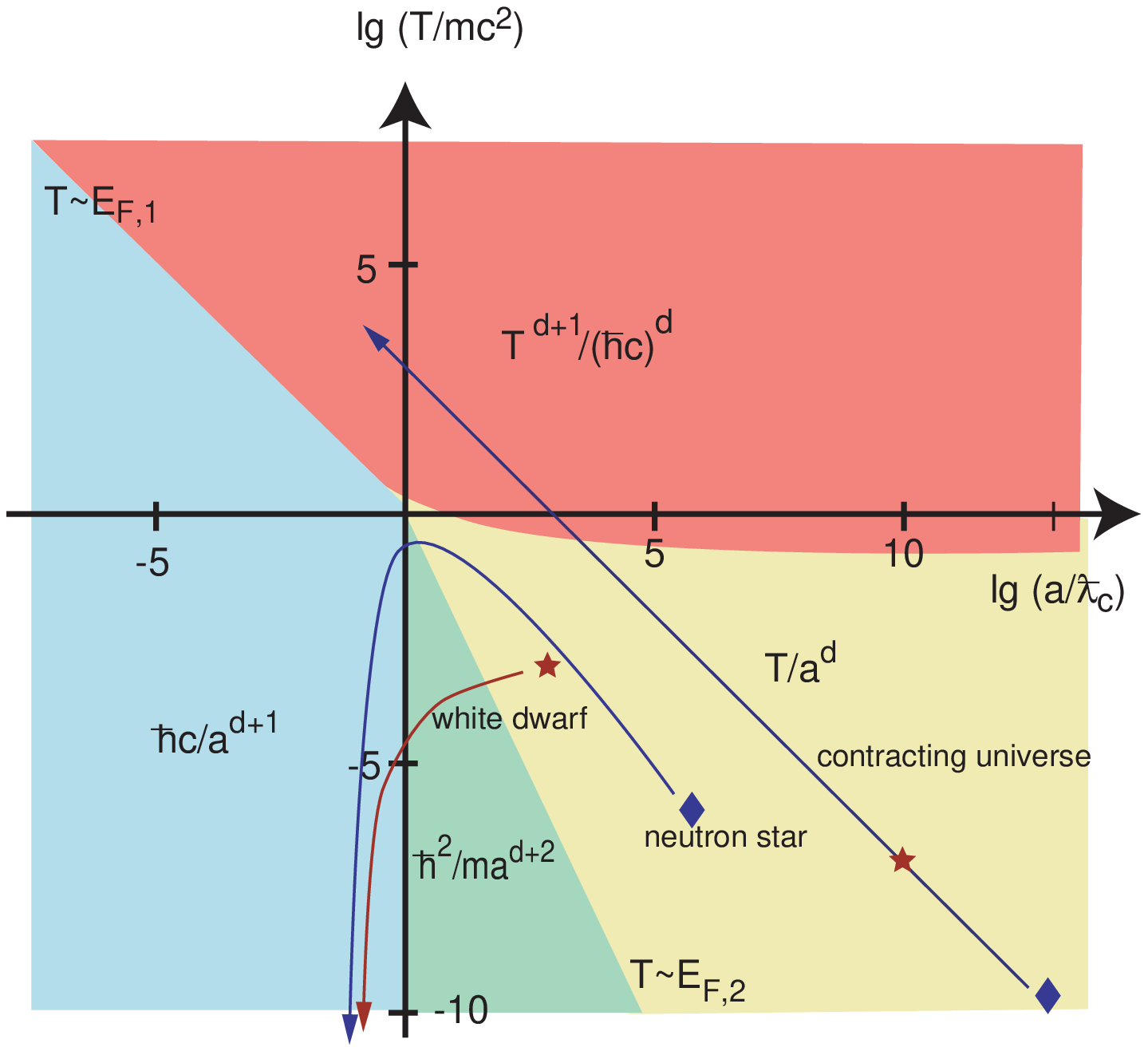}}
   \caption{\label{fig:ideal_gas1_fermions} The pressure
of an ideal  gas of massive fermions in different parameter regions.
The cross-over lines $T=E_{F,2}$ and $T=E_{F,1}$ separate the
    classical ($p\sim Ta^{-d}$) from the non-relativistic quantum
    region ($p\sim \hbar^2/(ma^{d+2})$
    and the radiation ($p\sim T^{d+1}/(\hbar c)^d$) from relativistic
    quantum region ($p\sim \hbar c/a^{d+1}$), respectively.
    $E_{F,\nu}(a)$ denotes the Fermi energy.
    The vertical  line $a=\lambdabar_C$ , $T<mc^2$
    separates the non-relativistic from the relativistic quantum
    behavior.
     The radiation region
 is characterized by pair creation processes where the
    {effective }
    particle distance is $\lambdabar_{T,1}<a$. The symbols
    $\blacklozenge$ and $\bigstar$ stand for the protons and electrons,
    respectively, which change their coordinates in this figure
    during the evolution of stars and the toy universe, respectively,  considered in
    section IV.
    }
   \end{figure}

\medskip

For \emph{very} \emph{dense} systems of atoms, where nuclei and
electrons form a  plasma with $a\ll a_{\text{Bohr}}$, electrons
and protons may \emph{recombine} to produce a neutron and a
neutrino (the inverse $\beta$-decay). The  charge of the resulting
nucleus is then reduced by one. In this case the electron density
 is no longer an independent variable but determined by the
chemical potential $\tilde\mu $ of this process. If   $T\ll
\tilde\mu$ and initially $a<\Lc$ the system is in the
ultra-relativistic region,   thermal fluctuations can be neglected.
$\tilde\mu$ determines then unambigously a new length scale
   \begin{equation}
   \lambdabar_{\mu}=\frac{\hbar c}{\tilde\mu}
   \label{eq:lambda}
   \end{equation}
corresponding to $\omega(\mu/T)\sim T/\mu$ in
(\ref{eq:lambda_grand_canonical}) which replaces $a$ in the region
$a<\lambdabar_{\mu}<\lambdabar_{T,1}$ in
eq.(\ref{eq:p_fermions_relativistic}). Since thermal fluctuations
are irrelevant and $a$ is not longer fixed, $T$ is replaced by
$\tilde \mu$ in all expressions. The pressure is therefore reduced
to \cite{ll1065}
   \begin{equation}
   p_F\sim\frac{{\tilde\mu}^{d+1}}{(\hbar c)^d}\,.
   \label{eq:p_inverse_beta_decay}
   \end{equation}

\medskip

\medskip

%%%%%%%             T    >>    m c^2        %%%%%%%%%%%%%%%%%%%%%%%
%%%%%

We come now to a discussion of the parameter region of ultra-high
temperatures  of fermionic systems where particle-antiparticle
\emph{pair production}  becomes important (regions (v) and (vi) of
Figure 1). In this region the particle number is not conserved and
it is more adequate to describe the system by specifying a chemical
potential, as we have done  in subsection
\ref{Grand_Canonical_Description}. For simplicity we assume that
only fermions of the same kind (and their corresponding
antiparticles) are generated. As discussed in Subsection
\ref{Quantum_case} the Coulomb interaction can be neglected since
the effective particle distance turns out to be much smaller than
the Bohr radius $a_B$. For temperatures $T\gg \mu $ the pressure is
determined by (\ref{eq:p_grand_canonical}) with $\omega(\mu/T)\sim
1$ and $\nu=1$ (since we are in the relativistic region), i.e.
\cite{ll617}
\begin{equation}\label{eq:p_fermions_radiation}
 p_{\text{F}}\sim\frac{T}{\lambdabar_{T,1}^d}
 \sim\frac{T^{d+1}}{(\hbar c)^d},
\end{equation}
 which agrees with the result for photons (compare
 (\ref{eq:p_grand_canonical})). Hence we
  will call (\ref{eq:p_fermions_radiation})
   \emph{radiation} \emph{behavior}.
  $\Lr$ has to be considered as
 the mean distance between the fermions which is here smaller than
 $a$ because of pair production. If several kinds of fermions exist
 each of it gives the same contribution (\ref{eq:p_fermions_radiation})
to the pressure.
\medskip

Finally we consider the cross-over  between
(\ref{eq:p_fermions_radiation}) and the pressure formulas valid at
lower  temperature.  We begin with the region $\Lc<a$. Lowering
temperature we  expect a cross-over to the classical result
(\ref{eq:p_classical}). To describe the cross-over we have to
understand the contribution of pair production if we raise
temperature starting in the classical region. It is therefore
convenient to use a grand canonical description also for the
description of this cross-over. To make contact with the classical
result (\ref{eq:p_classical}), we have to express the chemical
potential of particles and anti-particles in terms of the
parameters of the system. Particles and anti-particles
 can annihilate into photons. Since the sum of the chemical
potentials is conserved in this reaction \cite{ll1012} and the
chemical potential of photons is zero, we find that  the chemical
potentials of  particle and anti-particles have opposite sign, say
$\pm \mu$. From (\ref{eq:p_grand_canonical}) and
(\ref{eq:Omega_of_mu/T}) we conclude that in the classical region
the density of particles and anti-particles, respectively,   is
given by
\begin{equation}\label{eq:particle_antiparticle_density}
\frac{1}{\Ln^{d}}\,\exp\left({\frac{\,\,\pm \mu-mc^2}{ T}}\right).
\end{equation}
Here we have included explicitly the term $-mc^2$ resulting from the
shift of the energy scale discussed in \ref{Introduction}. If the
density of \emph{generated} particles is small compared to the
initial density $a^{-d}$ such that $ \Ln^{-d}e^{(\mu-mc^2 )/(
T)}\approx a^{-d}$ we get after elementary manipulations for the
density of the generated particles
\begin{equation}\label{eq:pair_density_fermions}
\frac{a^d}{\Ln^{2d}}\,\exp\left({\frac{-2mc^2}{T}}\right).
\end{equation}
The cross-over to the radiation region  sets in if the density
(\ref{eq:pair_density_fermions}) of  generated particles is a
finite fraction  of the initial density  $a^{-d}$.
% i.e. if $\Ln
%\lesssim a e^{-2mc^2/(dT)}$
This gives for the cross-over line $a_{co}(T)$ (compare Figure 2)
\begin{equation}\label{eq:fermions_crossover}
 \frac{a_{co}}{\Lc}\sim \sqrt{\frac{mc^2}{T}}
 \exp(\frac{mc^2}{Td}).
\end{equation}

 In the relativistic region $a<\Lc$ essentially the same
arguments apply but now
 we have to replace   $\Ln$ by $\Lr$ in
 (\ref{eq:particle_antiparticle_density}). The
 cross-over follows then at $\Lr\lesssim a\exp(-mc^2/Td)$ which
 corresponds for $a<\Lc$ essentially to $a\sim \Lr$ (compare
 Figure 2), in agreement with our previous conjecture.

%%%%%%%%%%%%%%%%%%%%%%%%%%%%%%%%%%%%%%%%%%%%%%%%%%%%%%%%%%%%%%%%%%%%%%

\subsection{\textsc{Strong} \textsc{quantum}\textsc{ limit - bosons}}

%%%%%%%%%%%%%%%%%%%%%%%%%%%%%%%%%%%%%%%%%%%%%%%%%%%%%%%%%%%%%%%%%%%%
%%%%%%

 For \emph{bosons} the  low temperature pressure cannot depend on
the volume since all bosons may accumulate in the state of lowest
energy. This requires in (\ref{eq:p_scaling_ansatz})
$\psi_B(x_{\nu})\sim x_{\nu}^d$ from which we get for the pressure
\begin{equation}\label{eq:p_bosons}
    p_{\text{B}}(\nu)\sim\frac{T}{\lambdabar_{T,\nu}^d }\sim
    \frac{T^{1+d/\nu}}{\hbar^d\gamma_{\nu}^{d/\nu}}
    \sim \left(\frac{T}{T_{c,\nu }}\right)^{d/\nu} \frac{T}{a^d},
    %\,\,\,\,\,\,
    %a<\lambdabar_{T,\nu}.
\end{equation}
where on the r.h.s. we introduced the temperature of \emph{Bose
condensation}
\begin{equation}\label{eq:Tc_Bose_condensation}
 T_{c,\nu}= T_{c,\nu}(a)\sim \frac{\hbar^{\nu}\gamma_{\nu}}{a^{\nu}}.
\end{equation}

In particular, we get for relativistic particles \cite{ll6016}
   \begin{equation}
   p_{\text{B}}(\nu=1)\sim\frac{T}{\lambdabar_{T,1}^d}\sim
   \frac{T^{d+1}}
   {(\hbar c)^d}
   \label{eq:p_bosons_relativistic}
   \end{equation}
and for non-relativistic particles \cite{ll599}
   \begin{equation}
   p_{\text{B}}(\nu=2)
   \sim\frac{T}{\lambdabar_{T,2}^d}
   \sim\frac{T^{(d+2)/2}}{\hbar^d}m^{d/2}\,.
   \label{eq:p_bosons_non-relativistic}
   \end{equation}
Again both the relativistic and the non-relativistic case can be
combined into the scaling form
   \begin{equation}
   p_{\text{B}}\sim\frac{T}{\lambdabar_{T,1}^{d}}
   \frac{\Psi_{\text{B}}(x_1,\frac{a}{\Lc})}{\psi_{\text{B}}(x_1)}\, .
   \label{eq:p_bosons_crossover}
   \end{equation}
The cross-over between (\ref{eq:p_bosons_non-relativistic}) and
(\ref{eq:p_bosons_relativistic}) occurs at $\Lr\approx \Lc$, i.e. at
$T\approx mc^2$.

Since $ a<\lambdabar_{T,\nu}$ the pressure is \emph{reduced} with
respect to the classical result (\ref{eq:p_classical}). This can be
explained by assuming that a finite fraction of particles sits now
in the state of zero energy and hence does not contribute to the
pressure. Comparing (\ref{eq:p_classical}) and (\ref{eq:p_bosons})
we conclude that $ \lambdabar_{T,\nu}$ has to be considered as the
mean distance of particles \emph{above} the condensate
\cite{condensate}.

   \begin{figure}[hbt]
  \centerline{\epsfxsize=9cm
   \epsfbox{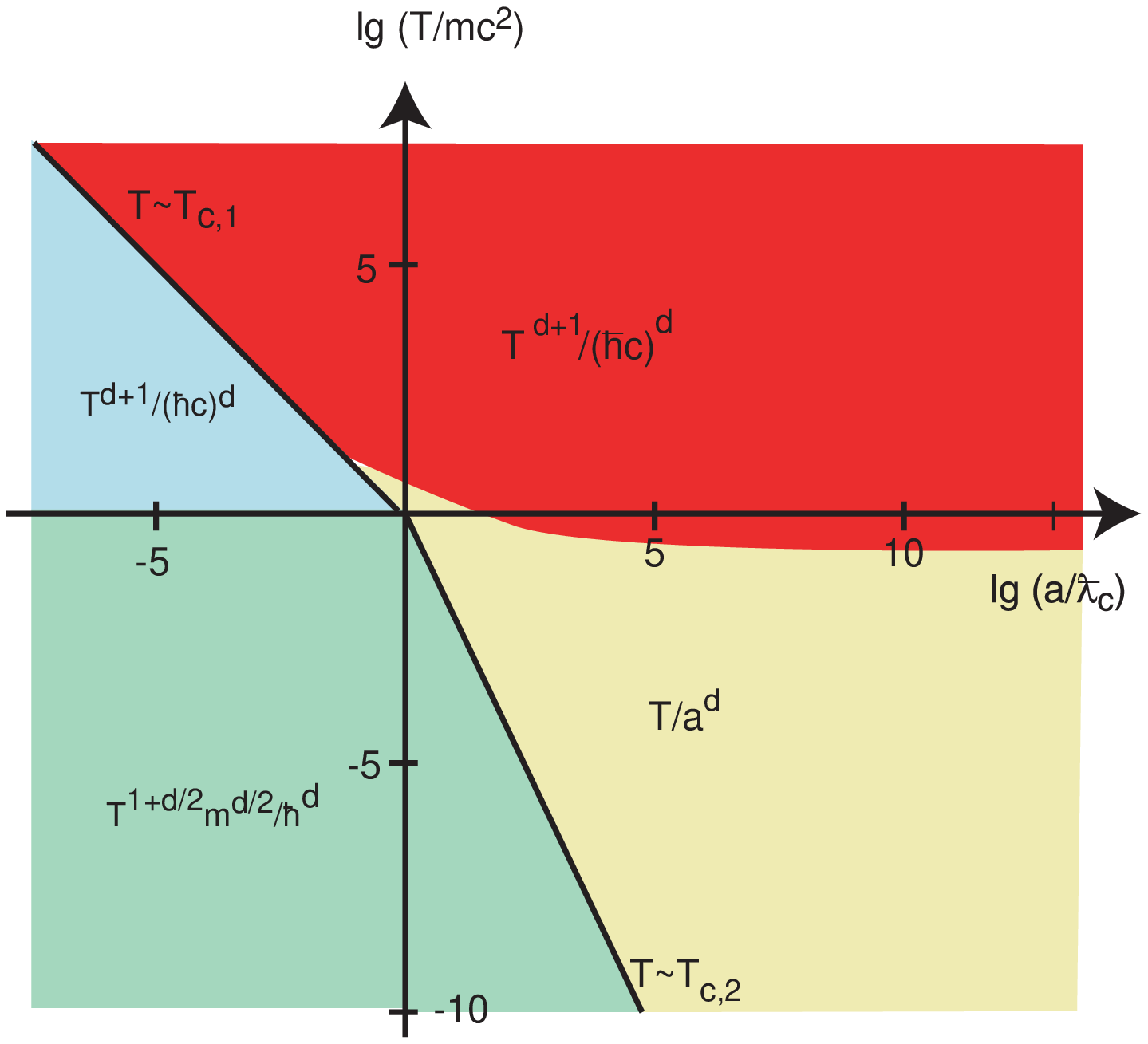}}
   \caption{\label{fig:ideal_gas1_bosons} The pressure of an ideal
   gas of massive bosons in different parameter regions, as discussed
   in the text.
   The line of Bose condensation  transition $T_{c,2}(a)$
   in the non-relativistic regime is continued in the relativistic
   regime as $T_{c,1}(a)$. There is an   additional
   cross-over at $T\sim mc^2$ from non-relativistic to relativistic
   behavior.
   }
   \end{figure}

Next we discuss  the phase diagram of massive Bose particles
obeying the full dispersion relation
(\ref{eq:relativistic_dispersion_relation}). We begin with the
classical region $\lambdabar_C<\Ln<a$ where the pressure is given
by (\ref{eq:p_classical}). % and $\frac{\hbar^2}{ma^2}<T\ll mc^2$
Reducing $T$ we reach at $a \approx \Ln$  the region where a
macroscopic fraction of particles does not contribute to the
pressure since it is in the state of zero energy. As argued above,
the number $N_{E>0}$ of  the remaining particles of finite energy
can be written as
\begin{equation}\label{}
   \frac{V}{\lambdabar_{T,2}^d}=
    N \left(\frac{a}{\lambdabar_{T,2}}\right)^d\sim
    N\left(\frac{T}{T_{c,2}}\right)^{d/2}=N_{E>0}.
\end{equation}
Thence the number of particles in the condensate is
\begin{equation}\label{eq:condensate}
    N_{0}=N- N_{E>0}=
    N\left(1-\left(\frac{T}{T_{c,2}}\right)^{d/2}\right),
\end{equation}
which shows that $T_{c,2}(a)$ is indeed the transition temperature
of Bose condensation \cite{llv1.62.4}. If one enters the condensed
phase by lowering $T$ at \emph{fixed} $a$, the total number of
bosons remains fixed whereas the mean distance of the particles
above the condensate is equal to $\Ln (T)$, which increases if we
further decrease temperature.

On the other
hand, if one approaches the transition line by decreasing $a$ at
\emph{fixed} $T$, the mean distance $\lambdabar_{T,2}$ between
particles above the condensate remains
 constant (since $T$ is constant) whereas the excess particles go
to the ground state. This behavior is seen also for
$a<\lambdabar_C$ since the effective distance of particles
contributing to the pressure remains
$\lambdabar_{T,2}(>\lambdabar_C)$
 (compare Fig.3). Thus, Bose condensation as described by equations
(\ref{eq:p_bosons_non-relativistic}) and (\ref{eq:condensate}) is
seen both in regions (ii) and  (iii) of Figure 1.

\medskip

At sufficiently high temperatures we have to expect that production
of particle-antiparticle  pairs sets in. For simplicity we will
again assume  that only bosons of the same type can be created (and
their anti-particles).  At temperatures $T\gg mc^2 $ bosons behave
relativistically and the
 pressure is given by (\ref{eq:p_bosons_relativistic}).

The discussion of the cross-over to (\ref{eq:p_classical}) if
$a>\Lc$ and to (\ref{eq:p_bosons_non-relativistic}) if $a<\Lc$,
 respectively,  follows
essentially the same lines as in the case of fermions, with the
important difference that in describing the cross-over in the region
$a<\lambdabar_{T,2}$ the effective particle distance is now $\Ln$
(instead of $a$ as in the case of fermions). From
(\ref{eq:pair_density_fermions}) we therefore conclude that the
cross-over occurs at $T\sim mc^2$. Note, that the cross-over is
smooth, since at $T\approx mc^2$, $\Ln\approx \Lr$ if $a<\Lc$.

Since in  the relativistic region (iv), $a<\Lr<\Lc$, the effective
particle density $\Lr^{-1}$ is smaller than $a^{-d}$ one is inclined
to conclude that here both pair creation and Bose condensation takes
place.
 For
 $a>\Lc$ the cross-over is the same as for fermions.

%%%%%%%%%%%%%%%%%%%%%%%%%%%%%%%%%%%%%%%%%%%%%%%%%%%%%%%%%%%%%%%%%%
%%%%%%%

\section{Illustrations }

%%%%%%%%%%%%%%%%%%%%%%%%%%%%%%%%%%%%%%%%%%%%%%%%%%%%%%%%%%%%%%%%%%%%%%

\subsection{Short history of a star}

An important application of the various equations of state can be
found in the history of \emph{stars} which we present  here in a
cartoon like picture. Initially the star burns hydrogen to helium
and other light elements and keeps in this way its temperature
$T$. The ratio $T/(m_pc^2)$ is of the order $10^{-6}$, where $m_p$
is the proton mass. This is the present situation of our sun.
 The pressure of a
star can roughly be described by the classical equation
(\ref{eq:p_classical}), which counteracts the gravitational
pressure
\begin{equation}\label{eq:star}
  p_{\,G}%\sim \frac{m_pc^2}{a^3}\frac{R_S}{R}
  \sim G_N\frac{m_p M}{a^3R}\sim G_N\frac{m_p^{4/3}M^{2/3}}{a^4},
\end{equation}
 Here $R$  and $M$
 denote the radius and the mass
 of the star, respectively, and $G_N$ is the
 gravitational constant.
We do not present  the derivation of  (\ref{eq:star}), which is
straightforward, but remark that $p_G\cdot R^3$ is the
gravitational energy of a sphere $\sim G_N M^2/R$, as one obtains
also from a dimensional argument. The ratio of the mass to the
radius of a star in this state is controlled by the thermal
equation of state (\ref{eq:p_classical}). Indeed, from the
equality of (\ref{eq:p_classical}) and (\ref{eq:star}) follows
$M/R\sim  T/(G_N m_p)$, which increases linearly with temperature.

 Once the nuclear fuel is exhausted, temperature cannot be
kept and the star collapses under the graviational force. As a
result the density increases and the star matter crosses over to the
region (ii) of Figure 1, where the pressure is given by
(\ref{eq:p_fermions_non-relativistic}). Since the pressure is now
inversely proportional to the mass, mainly the electrons contribute
to it. Once under further contraction $a$ has  reached $\Lc$
(corresponding to densities of the order $\sim 10^7
\text{gcm}^{-3}$) (\ref{eq:p_fermions_relativistic}) has to be
applied for the pressure. The star has reached the state of a
\emph{white dwarf}(compare Figure 2).

If the density is increased further ( i.e. $a\lesssim
\lambdabar_{\mu}$), neutrons are generated via the inverse
$\beta$-decay.  The pressure is determined then by
(\ref{eq:p_inverse_beta_decay}) until most of the nucleons became
neutrons. For the pressure in \emph{neutron stars} formulas
(\ref{eq:p_fermions_non-relativistic}) for $a>\Lc$ and
(\ref{eq:p_fermions_relativistic}) for $a<\Lc$ again apply,  but the
mass $m_n\approx m_p$ is now that of the neutrons (compare Figure
2). This is the fate of stars with a mass between 15 and 30 times
the mass of the sun.

In the ultra-relativistic regions of white dwarfs and neutrons
stars the equality between (\ref{eq:star}) and
(\ref{eq:p_fermions_relativistic}) results in a relation for the
mass
\begin{equation}\label{eq:chandrasekhar_mass}
  M_C=m_p\left(\frac{\hbar c}{G_N m_p^2}\right)^{3/2}.
\end{equation}
This so-called \emph{Chandrasekhar mass} $M_C$ is the maximum mass
of white dwarfs and neutron stars. Its value is determined by the
thermal equation of state in region (iii) and is of the order of
twice the mass of the sun.

\subsection{A toy universe}

As a second illustration let us consider a \emph{contracting toy
universe} . For simplicity, this universe is supposed to have no
gravity, i.e. there are no stars, clouds of cosmic dust etc., just
photons and diluted massive particles. To make things even simpler
the only particles assumed to exist are hydrogen atoms ( a
''non-metallic'' universe in the language of astronomers). Initially
the photon spectrum is assumed to correspond to black body radiation
of about 3\,K. Since for massless particles $\lambdabar_{T,1}$ is
the only existing length scale, the typical distance of photons is
of the order $\lambdabar_{T,1}\sim 10^{-3}\,{\rm m}$.

\medskip

The number of H-atoms per ${\rm m}^3$ is assumed to be of  order 1.
Hydrogen atoms are bosons and hence could be represented in Figure
\ref{fig:ideal_gas1_bosons}. The horizontal and vertical coordinates
of the hydrogen gas in Figure \ref{fig:ideal_gas1_bosons} are given
by ${a_H}/{\lambdabar_{C,H}}\sim 10^{16}$ and $T/mc^2 \approx
10^{-16}$, respectively, i.e. they are {outside} of the  margins of
this figure. $\lambdabar_{C,H}\sim 2.1\,\,\cdot 10^{-16}\,{\rm m}$
is the Compton wave length of hydrogen.

\medskip

 Next we switch on the cosmic contraction, i.e. we go
backwards in time. It is convenient to introduce the contraction
factor $r(t)=a(t)\big/a(t_f)<1$, where $a(t)$ denotes the distance
at an earlier time $t<t_f$ when the final distance was
$a(t_f)\equiv a$. Since contraction of space reduces the
wavelength of photons such that
$\lambdabar_{T,1}(t)=\lambdabar_{T,1}(t_f)r(t)\equiv{\hbar
c}\big/{T_{\gamma}(t)}$ with $T_{\gamma}(t)=
T_{\gamma}(t_f)\big/r(t)$, i.e. the photon gas heats up.

\medskip

Without coupling to the photons, the hydrogen atoms form a classical
non-relativistic gas (region (i) of Figure 1) which contracts
adiabatically. Its temperature $T_H(t)=T_H(t_f)/r^2(t)$ rises
proportional to $r^{-2}(t)$, \emph{provided} there is enough
interaction between the hydrogen atoms to reach thermal equilibrium.
The horizontal and vertical coordinates of the hydrogen gas in
Figure 3 thence change  according to
${a_H(t)}/{\lambdabar_{C,p}}\sim r(t)$ and
 $T(t)/mc^2\sim r^{-2}(t)$, respectively.

\medskip

When the temperature of the photons reaches about $3000\,{\rm K}$
($r(t)\sim 10^{-3}$), the hydrogen atoms have a distance of about
$a_H(t)\sim 10^{-3}\,{\rm m}$ and are now partially ionized, i.e. we
have free protons and electrons, both fermions, which appear in
Figure \ref{fig:ideal_gas1_fermions}. Since both particles are
charged, they will emit and adsorb photons. From now on matter and
radiation have approximately the same temperature $T(t)\sim 3
\,r^{-1}(t) K$. The initial coordinates are
 $a_p(t)/\lambdabar_{C,p} \sim 10^{13}$,
 $T/m_pc^2=({\Lc}_{,p}/\lambdabar_{T,\nu})^{\nu} \sim 10^{-10}$
( i.e. $a_p(t)/{\Ln}_p\sim 10^8$) for protons and
$a_e(t)/\lambdabar_{C,e} \sim 10^{10}$, $T/m_ec^2
=({\Lc}_{,e}/\lambdabar_{T,\nu})^{\nu}\sim 10^{-7}$ ( i.e.
$a_e(t)/{\Ln}_e\sim 10^6$) for electrons, respectively. Thus their
initial positions are in region (i) of Figure 1. As long as $a_T/\Lc
\sim 10^4\ll a(t)/\Lc$, electrons and protons form initially an
non-interacting {non}-{relativistic} {classical} plasma (compare
Figure 2).
 Since $a(t)\sim r(T)$ and
$T(t) \sim r^{-1}(t)$, the slope of the path under further
contraction  in the double logarithmic plot of Figure 2 is $-1$
and remains unchanged  until we reach the relativistic region
$T>mc^2$. \medskip

 When the contraction parameter reaches $r(t)\sim
10^{-10}$, i.e temperatures reach about $3\cdot 10^{10}K$, the
mean particle distance is of the order of the Bohr radius.

When temperatures reach  the rest energy of the electrons, i.e.
crossing the bisecting  line $T=m_ec^2$ of Figure 2, the electrons
start to behave relativistically.  For $T\gg m_ec^2$
electron-positron pairs are formed and leptons behave
ultra-relativistically. For temperatures $T>10^{15} K$ and $a_p\sim
10^{-15}m$ hadrons decay into quarks and anti-quarks, which also
behave ultra-relativistically. Also their mean distance is now
determined by pair production. In addition   gluons, i.e. the bosons
which carry the interaction between quarks, have to be considered.
Eventually the coordinates of all fundamental fermions and bosons
reach the line $a\sim \Lr$.

The pressure in all cases is that or ultra-relativistic fermions
(\ref{eq:p_fermions_radiation}) or bosons
(\ref{eq:p_bosons_relativistic}), respectively. The total pressure
is the sum of these contributions, i.e. it is proportional to the
number of different fermions and bosons.

%%%%%%%%%%%%%%%%%%%%%%%%%%%%%%%%%%%%%%%%%%%%%%%%%%%%%%%%%%%%%%%%%
%%%%%%%%%

\section{Additional thermodynamic relations}

%%%%%%%%%%%%%%%%%%%%%%%%%%%%%%%%%%%%%%%%%%%%%%%%%%%%%%%%%%%%%%%%%%
%%%%%%%%%

 We can also use the results of the previous Chapters
 to determine  the {scaling} {behavior} of the \emph{free}
\emph{energy} $f(T,v)$ per particle which can be written as
 \begin{equation}
   f(T,v)=-T\varphi_{\text{F/B}}(x_{\nu})\,.
   \label{eq:f}
   \end{equation}
Differentiation with respect to $v=a^d$ gives
   \begin{equation}
   -\frac{\partial f}{\partial v}=\frac{T\varphi_{\text{F/B}}^{\prime}(x_{\nu})x_{\nu}}{vd}
   =p\,.
   \label{eq:dif_f}
   \end{equation}
Thus
$\psi_{\text{F/B}}(x_{\nu})=(1/d)x_{\nu}\varphi_{\text{F/B}}^{\prime}(x_{\nu})$.
The entropy per particle $s$ follows from
\begin{equation}
   s=-\frac{\partial f}{\partial T}=
   \varphi_{\text{F/B}}(x_{\nu})+\frac{1}{\nu}\varphi_{\text{F/B}}^{\prime}(x_{\nu})x_{\nu}
\end{equation}
and hence we get for the energy per particle $u$
   \begin{equation}
   u=f+Ts=\frac{T}{\nu}\varphi_{\text{F/B}}^{\prime}(x_{\nu})x_{\nu}=\frac{d}{\nu}pv
   \end{equation}
which is the well known relation between the energy $u$ and $pv$
both for relativistic and non-relativistic particle in any
dimension\cite{ll558}. For the chemical potential $\mu$ we obtain
from the Gibbs-Duham relation
   \begin{equation}
   \mu={f+pv}{}=T\left(\frac{1}{d}\varphi_{\text{F/B}}^{\prime}(x_{\nu})x_{\nu}-\varphi_{\text{F/B}}(x_{\nu})
   \right)\,.
   \label{eq_mu}
   \end{equation}
With our Ans\"atze used above we find at low temperatures $s=0$ for
fermions and $\mu=0$ for bosons.

\medskip

These expressions can be used to calculate additional thermodynamic
quantities. The \emph{specific} \emph{heat} follows from
(\ref{eq:dif_f}) as
   \begin{equation}
   c_T=\frac{\partial u}{\partial T}=\frac{d}{\nu}
   \frac{\partial}{\partial T}(pv)=\frac{d}{\nu}
   \frac{\partial}{\partial T}
   \big(T\psi_{\text{F/B}}(x_{\nu})\big)\,.
   \label{eq:specif_heat}
   \end{equation}
For \emph{fermions} our Ansatz $\psi_F(x_{\nu})\sim x^{-\nu}$, i.e.
$f\sim E_F$, gives $c_T=0$ instead of the correct result $c_T\sim
{T}/{E_F}\sim x_{\nu}^{\nu}$. This is not  surprising since we
determined $v p\sim f$ from the condition that the free energy
should go to a constant (the Fermi energy) for $T\rightarrow 0$.

To get the correct specific heat we have to add finite temperature
corrections to $u$. Indeed, the low temperature corrections to
energy of the \emph{Fermi gas} result from the difference $\Delta$
of the Fermi distribution at non-zero and zero temperatures (the
latter is the step function). These corrections are of the form
\cite{sommerfeld}
   \begin{equation}
   u=\int\limits_{-\infty}^{\infty}dE\,g(E)\,{
   E}\,
   \Delta\left(\frac{E-\mu}{T}\right)
   \label{eq:I}
   \end{equation}
where $\Delta(x)=
   (e^x+1)^{-1}-
   \theta(-x)\,$,
and $g(E)$ is the density of states which is smooth at $E=\mu$.
With $\Delta(x)=-\Delta(-x)$ it is clear that the low temperature
expansion of $u$ includes only \emph{even} powers in $T$. Since
$E_F$ is the only energy scale, the low temperature expansion of
the energy of the ideal Fermi gas is therefore of the form
\begin{equation}\label{}
u\sim E_F(1+\frac{T^2}{E_F^2}+c_2\frac{T^4}{E_F^4}+\cdots   )
\end{equation}
 which is the Sommerfeld expansion. This gives  for the specific
heat $c_T\sim T/E_F$ \cite{ll576}. On the other hand, calculating
the compressibility we get with $\psi_F(x_{\nu})\sim x_{\nu}^{-\nu}$
immediately
  the correct result
\begin{equation}
   \kappa=-\frac{1}{V}\frac{\partial V}{\partial p}=
   \frac{d}{\nu+d}p^{-1}.
   \label{eq:kappa}
   \end{equation}

\medskip

 {For} \emph{bosons} we obtain with $\psi_B\sim x^d$ the
correct result for the specific heat \cite{ll626,ll6315}
$c_T\sim\frac{\partial}{\partial T}
(Tx_{\nu}^d)\sim\left({T}/{T_c}\right)^{d/\nu}\sim\left({a}/
{\lambdabar_{T,\nu}} \right)^d$ and  the compressibility diverges.

\medskip

To conclude we have shown that the main  properties of ideal quantum
gases can be obtained from simple dimensional arguments and the
Pauli principle, without resorting to statistical mechanics.

%%%%%%%%%%%%%%%%%%%%%%%%%%%%%%%%%%%%%%%%%%%%%%%%%%%%%%%%%%%%%%%%%%%
%%%%
%%%

\bigskip

\textsc{\textbf{Acknowledgment}}

%%%%%%%%%%%%%%%%%%%%%%%%%%%%%%%%%%%%%%%%%%%%%%%%%%%%%%%%%%%%%%%%%%%%
%%%%%%%

It is a pleasure to acknowledge helpful advice on the preparation of
this paper by  A. Glatz, C. Kiefer, J. Krug, E. M\"uller-Hartmann,
A. Rosch, B. Rosenow, S. Scheidl, and D. Stauffer.

%%%%%%%%%%%%%%%%%%%%%%%%%%%%%%%%%%%%%%%%%%%%%%%%%%%%%%%%%%%%%%%%%%%%%
%%%


\begin{thebibliography}{99}
\bibitem{notnew} Most likely, some of the arguments presented in
 this paper are not new, but I have
never encountered this formulation elsewhere.
\bibitem{quasiparticles} Strictly speaking also
quasi-particles exhibit a dispersion which we ignore here.
\bibitem{dimensional_analysis} To get this relation one starts from
the Ansatz for the pressure $p\sim
T^{\alpha}a^{\beta}\gamma_{\nu}^{\delta}$ and compares the powers
of the dimensions of $M, L, t$ on both sides of this relations,
which determines the exponents $\alpha, \beta, \delta$.
\bibitem{Boltzmann_constant} Thus we put $k_B=1 $, 1
 Kelvin corresponds roughly to $10^{-4}\,{\rm eV}$.
\bibitem{powers} Trivially, any power of $x_{\nu}$ is also a
solution.
\bibitem{lambda} Note that in the classical
region $\Lc<\lambdabar_{T,\nu}<a$ the relevant thermal de Broglie
length is $\Ln$ whereas in the ultra-relativistic quantum region
$a<\lambdabar_{T,\nu}<\Lc$  this role is played by $\Lr$.
\bibitem{LLV1_56_9} These results can be found, including
the numerical prefactors, which remain undetermined in our approach,
e.g. in Reference \cite{LLV1}, eq. (56.9)
\bibitem{probability} Considering the partition of two particles on
two
states $|x\rangle$, $|y\rangle$ of equal energy, classically there
are 4 possibilities, $|x\rangle|x\rangle$, $|y\rangle|y\rangle$,
$|x\rangle|y\rangle$ and $|y\rangle|x\rangle$. The probability that
two particles are in the same state, say $|x\rangle$, is 1/4. In
quantum mechanics only the states $|x\rangle|x\rangle$,
$|y\rangle|y\rangle$ and
$\frac{1}{\sqrt{2}}\big(|x\rangle|y\rangle+|y\rangle|x\rangle\big)$
exist, i.e. the probability for having two particles in the same
state is 1/3.
\bibitem{condensate} For particles with vanishing chemical potential like photons or
quasi-particles there is no particle number conservation and hence
no condensate, particles just disappear from the system.
\bibitem{llv1.62.4} Reference \cite{LLV1}, eq. (62.4). We have
chosen here the numerical prefactor of $T_{c,2}$, which remains
undetermined in our approach, such that a non-zero condensate
appears first at $T_{c,2}$.
\bibitem{bh}  A star of
radius $R\le R_S$ would be a black hole.
\bibitem{sommerfeld}  Here we use a modest amount of statistical
mechanics.
\end{thebibliography}
\end{document}